\documentclass[pra,aps,twocolumn,eqsecnum,superscriptaddress,showpacs,amsmath]{revtex4-1}
\usepackage{amsmath,amssymb,graphicx,graphics,color}
\usepackage{dcolumn}% Align table columns on decimal point
\usepackage{bm}% bold math
\usepackage{psfrag}
\usepackage{graphicx}
\usepackage{latexsym}
\usepackage{epsfig}
\usepackage{color}
\usepackage{bm}% bold math
\usepackage{amsmath}
\usepackage{subfigure}

\begin{document}

\title{Pressure-induced collapse of spin-orbital Mott state in the hyperhoneycomb iridate $\beta$-Li$_2$IrO$_3$}
\author{T. Takayama}
\affiliation{Max Planck Institute for Solid State Research, Heisenbergstrasse
  1, 70569 Stuttgart, Germany}
\affiliation{Institute for Functional Matter and Quantum Technologies, University of Stuttgart, Pfaffenwaldring 57, 70550 Stuttgart, Germany}
\author{A. Krajewska}
\affiliation{Max Planck Institute for Solid State Research, Heisenbergstrasse
  1, 70569 Stuttgart, Germany}
\affiliation{Institute for Functional Matter and Quantum Technologies, University of Stuttgart, Pfaffenwaldring 57, 70550 Stuttgart, Germany}
\author{A. S. Gibbs} 
\affiliation{ISIS Facility, STFC Rutherford Appleton Laboratory, Chilton, Didcot, Oxon OX11 0QX, UK}
\author{A. N. Yaresko}
\affiliation{Max Planck Institute for Solid State Research, Heisenbergstrasse
  1, 70569 Stuttgart, Germany}
\author{H. Ishii}
\affiliation{National Synchrotron Radiation Research Center, Hsinchu 30076, Taiwan}
\author{H. Yamaoka}
\affiliation{RIKEN SPring-8 Center, Sayo, Hyogo 679-5148, Japan}
\author{K. Ishii}
\affiliation{Synchrotron Radiation Research Center, National Institutes for Quantum and Radiological Science and Technology, Sayo, Hyogo 679-5148, Japan}
\author{N. Hiraoka}
\affiliation{National Synchrotron Radiation Research Center, Hsinchu 30076, Taiwan}
\author{N. P. Funnell}
\affiliation{ISIS Facility, STFC Rutherford Appleton Laboratory, Chilton, Didcot, Oxon OX11 0QX, UK}
\author{C. L. Bull}
\affiliation{ISIS Facility, STFC Rutherford Appleton Laboratory, Chilton, Didcot, Oxon OX11 0QX, UK}
\author{H. Takagi}
\affiliation{Max Planck Institute for Solid State Research, Heisenbergstrasse
  1, 70569 Stuttgart, Germany}
\affiliation{Institute for Functional Matter and Quantum Technologies, University of Stuttgart, Pfaffenwaldring 57, 70550 Stuttgart, Germany}
\affiliation{Department of Physics,
  University of Tokyo, 7-3-1 Hongo, Tokyo 113-0033, Japan}

\date{\today}
\begin{abstract}
Hyperhoneycomb iridate $\beta$-Li$_2$IrO$_3$ is a three-dimensional analogue of two-dimensional honeycomb iridates, such as $\alpha$-Li$_2$IrO$_3$, which recently appeared as another playground for the physics of Kitaev-type spin liquid. $\beta$-Li$_2$IrO$_3$ shows a non-collinear spiral ordering of spin-orbital-entangled $J_{\rm eff}$ = 1/2 moments at low temperature, which is known to be suppressed under a pressure of $\sim$2 GPa. With further increase of pressure, a structural transition is observed at $P_{\rm S}$ $\sim$ 4 GPa at room temperature. Using the neutron powder diffraction technique, the crystal structure in the high-pressure phase of $\beta$-Li$_2$IrO$_3$ above $P_{\rm S}$ was refined, which indicates the formation of Ir$_2$ dimers on the zig-zag chains, with the Ir-Ir distance even shorter than that of metallic Ir. We argue that the strong dimerization stabilizes the bonding molecular orbital state comprising the two local $d_{zx}$-orbitals on the Ir-O$_2$-Ir bond plane, which conflicts with the equal superposition of $d_{xy}$-, $d_{yz}$- and $d_{zx}$- orbitals in the $J_{\rm eff}$ = 1/2 wave function produced by strong spin-orbit coupling. The results of resonant inelastic x-ray scattering (RIXS) measurements and the electronic structure calculations are fully consistent with the collapse of the $J_{\rm eff}$ = 1/2 state. A subtle competition of various electronic phases is universal in honeycomb-based Kitaev materials.
\end{abstract}

%\keywords{}

\maketitle
\section{Introduction}

The Kitaev model, with $S$ = 1/2 spins on a honeycomb lattice connected by a bond-dependent Ising coupling, has been attracting considerable interest as it provides an exactly solvable quantum-spin-liquid (QSL) ground state consisting of the two kinds of Majorana fermions \cite{Kitaev_AP}. The materialization of the Kitaev QSL has been pursued extensively in honeycomb-based spin-orbital Mott insulators with heavy transition metal ions with $d^5$($t_{2g}^5$) configuration, such as Ir$^{4+}$ and Ru$^{3+}$ \cite{Kee_review, Winter_review}. In these spin-orbital compounds, the heavy $d^5$ transition metal ions are octahedrally coordinated with anions and the octahedra form a honeycomb network by sharing their edges. The strong spin-orbit coupling $\lambda_{\rm SO}$ $\sim$ 0.5 eV for Ir$^{4+}$ and 0.1 eV for Ru$^{3+}$ splits the degenerate $t_{2g}^5$ into the half-filled $J_{\rm eff}$ = 1/2 doublet and the completely filled $J_{\rm eff}$ = 3/2 quartet \cite{BJ_Science2009}. The magnetism of the candidate compounds therefore originates from $J_{\rm eff}$ = 1/2 pseudo-spins. The super-exchange coupling between two adjacent $J_{\rm eff}$ = 1/2 moments is shown to be a bond-dependent ferromagnetic Ising interaction as in the Kitaev model \cite{Jackeli_PRL2009}.

The layered honeycomb iridates Na$_2$IrO$_3$ and $\alpha$-Li$_2$IrO$_3$ emerged as the first generation of candidate materials for the Kitaev QSL \cite{Singh_Na2IrO3, Singh_Li2IrO3}. Their ground state, however, turned out not to be a QSL. They were found to show a magnetic transition to a zigzag-type antiferromagnetic phase \cite{Na2IrO3_RXS, Na2IrO3_neutron} and to a non-coplanar spiral phase \cite{alpha_RXS} respectively at a low temperature. $\alpha$-RuCl$_3$ was subsequently proposed as the first non-iridium-based candidate but again was found to show a zigzag-type antiferromagnetic ordering as in Na$_2$IrO$_3$ \cite{Plumb_RuCl3, Johnson_RuCl3, Cao_RuCl3, RuCl3_Nmat}. In parallel with this, three-dimensional (3D) analogues of the two-dimensional (2D) honeycomb $\alpha$-Li$_2$IrO$_3$, $\beta$-Li$_2$IrO$_3$ and $\gamma$-Li$_2$IrO$_3$, were discovered as another platform for Kitaev magnetism \cite{beta-Li2IrO3, gamma-Li2IrO3}. These 3D honeycomb compounds also show a clear magnetic transition into a complex spiral phase \cite{beta-Li2IrO3_RXS, gamma-Li2IrO3_RXS} similar to that of $\alpha$-Li$_2$IrO$_3$, though a closer proximity to the Kitaev spin liquid than their 2D analogues is suggested \cite{beta-Li2IrO3_EPL, beta-Li2IrO3_QC}, for example, by the ferromagnetic Curie-Weiss temperature. The presence of magnetic interactions other than the Kitaev coupling, such as a direct Heisenberg exchange and off-diagonal coupling, has been discussed to stabilize the long-range magnetic ordering instead of a QSL state \cite{Kee_review, Winter_review}.

In the 2D honeycomb iridates, a chemical substitution of the interlayer Li ions was attempted to tune the magnetic interactions through a local lattice distortion and to bring the ground state closer to the QSL \cite{Ag3LiIr2O6, Cu3LiIr2O6, Cu2IrO3}. With such an approach, H$_3$LiIr$_2$O$_6$ was very recently found to host a QSL ground state \cite{H3LiIr2O6, H3LiIr2O6_structure}, while the relevance to Kitaev physics remains yet to be identified. Control of the ground states using magnetic field was also attempted. By applying a magnetic field of $\mu_{0}H_{c}$ $\sim$ 8 T parallel to the honeycomb planes in $\alpha$-RuCl$_3$, a disappearance of the magnetic ordering was observed. The emergence of a QSL-like phase was pointed out and has been a subject of intensive studies \cite{Johnson_RuCl3, RuCl3_C, RuCl3_thermal_conductivity, RuCl3_NMR, RuCl3_Kasahara, RuCl3_theory}.

Another promising approach to control the magnetic ground state may be the application of pressure. The emergence of a QSL state under pressure was theoretically proposed in honeycomb-based iridates \cite{pressure_theory, beta-Li2IrO3_Kee, beta-Li2IrO3_Perkins}. Indeed, the suppression of long-range magnetic order under high pressure was reported in 3D honeycomb iridates $\beta$-Li$_2$IrO$_3$ and $\gamma$-Li$_2$IrO$_3$, in x-ray magnetic circular dichroism \cite{beta-Li2IrO3, beta-Li2IrO3_Veiga}, resonant magnetic x-ray scattering \cite{gamma-Li2IrO3_pressure}, and muon spin rotation measurements \cite{beta-Li2IrO3_Gegenwart}. In $\beta$-Li$_2$IrO$_3$, possibly related to the disappearance of the magnetic order at low temperatures, a first order pressure-induced structural transition at room temperature and at critical pressure $P_{\rm S}$ $\sim$ 4 GPa was discovered, accompanied by a discontinuous jump of the lattice constants \cite{beta-Li2IrO3_Veiga}. The impact of this structural phase transition on the electronic structure remains yet to be clarified.

In this paper, we show that the structural transition at $P_{\rm S}$ $\sim$4 GPa in $\beta$-Li$_2$IrO$_3$ is accompanied by the formation of Ir$_2$ dimers on the one-dimensional zig-zag chains. Through resonant inelastic x-ray scattering and electronic structure calculations, we show that the dimer consists of local molecular orbitals derived from Ir $t_{2g}$ electrons, indicating a breakdown of the $J_{\rm eff}$ = 1/2 description in the high-pressure phase. The formation of dimers may give a clue to the origin of the putative QSL behavior appearing prior to the structural transition, and points to a subtle balance of various competing electronic phases in the honeycomb-based iridates.

\section{Experimental}

Neutron diffraction measurements on a powder sample of $\beta$-Li$_2$IrO$_3$ were performed under pressure to unveil the detailed crystal structure of the high pressure phase.  The use of neutron diffraction allowed the reliable and precise refinement of not only the heavy Ir positions but also those of light Li and O to which neutrons are intrinsically much more sensitive than x-rays. The measurements were conducted at the Pearl beamline of the ISIS neutron source \cite{Pearl}. Pressure was applied by a Paris-Edinburgh press up to 5.2 GPa \cite{PE_press}. The anvils were single-toroidal zirconia-toughened alumina (ZTA), and an encapsulated TiZr gasket was used \cite{TiZr_gasket}. In order to minimize the neutron absorption by $^{191}$Ir, we prepared an isotope-enriched powder sample of $\beta$-$^7$Li$_2$$^{193}$IrO$_3$ \cite{Ir_isotope}. Deuterium substituted methanol-ethanol mixture (4:1 by volume) was used as a hydrostatic pressure medium.  The applied pressure was calibrated from the lattice constant of NaCl powder added as a pressure-marker. All of the measurements were conducted at room temperature. The Rietveld refinement of diffraction patterns was performed by assuming the presence of five phases, $\beta$-Li$_2$IrO$_3$, metallic Ir as an impurity, NaCl and anvil materials (ZrO$_2$ and Al$_2$O$_3$), using the GSAS program \cite{GSAS}. 

To investigate the electronic structure under pressure, we performed resonant inelastic x-ray scattering (RIXS) measurements with Ir $L_3$-edge on $\beta$-Li$_2$IrO$_3$ at BL12XU of SPring-8. A diamond anvil cell (DAC) was used for the application of pressure. A small single crystal of 50 $\mu$m size, grown by a flux method, was loaded in a DAC with Fluorinert (1:1 mixture of FC-70 and FC-77 by volume) as a pressure medium \cite{Supplementary}. Pressure was evaluated by the fluorescence spectra of a ruby ball loaded together with the sample. A gasket made of beryllium was used so that the incident and scattered x-rays go through the gasket with minimum attenuation. The energy of incident x-rays was tuned to 11.215 keV, which corresponds to Ir $2p_{3/2}$ – 5$d$($t_{2g}$) excitation. The incident x-ray beam was monochromated by a Si(111) double-crystal monochromator and further by a 4-bounced Si(440) high-resolution monochromator, and focused by using a Kirkpatrick-Baez mirror. The scattered x-rays were analyzed by a diced and spherically-bent Si(844) analyzer. The total energy resolution, estimated from the full width at half-maximum of the elastic line, was about 100 meV. No $\bm{q}$-resolved measurements were performed, and the obtained spectra are regarded as $\bm{q}$-averaged ones.  For a reference, we also collected the RIXS spectrum of polycrystalline $\beta$-Li$_2$IrO$_3$ at ambient pressure without using a DAC. All data were collected at room temperature.  

The electronic structure calculations were performed with the crystal structures refined from neutron diffraction data. The calculations were carried out based on the local density approximation (LDA) using the fully relativistic linear muffin-tin orbital (LMTO) method implemented in the PY LMTO code \cite{calculation_ref}. Spin-orbit coupling was taken into account by solving the four-component Dirac equation inside an atomic sphere. This allows to obtain $J$ resolved densities of states.

\section{Results}
\subsection{Structural transition under pressure}

The result of the structure refinement at ambient pressure from neutron diffraction data is shown in Supplemental Materials Fig. S1(a) and Table S1 \cite{Supplementary}, which agrees very well with that obtained by single-crystal x-ray diffraction [15]. $\beta$-Li$_2$IrO$_3$ crystallizes in an ordered rock-salt-type structure. Each IrO$_6$ octahedron shares its edges with the three neighboring IrO$_6$ octahedra as in $\alpha$-Li$_2$IrO$_3$ \cite{beta-Li2IrO3}. The local configuration of bonds around an IrO$_6$ octahedron is illustrated in Fig. 1(a). All of the IrO$_6$ octahedra are crystallographically equivalent and form a three-dimensional network via the three almost 120$^{\circ}$ bonds, termed a hyperhoneycomb lattice. The sub-lattice of Ir atoms at room temperature and at ambient pressure is depicted in Fig. 1 (b). The hyperhoneycomb lattice can be viewed as an assembly of Ir zig-zag chains, running along the $\bm{a}+\bm{b}$ and the $\bm{a}-\bm{b}$ directions alternately. The zig-zag chains are bridged by the Ir-Ir bonds along the $c$-axis. The three 120$^{\circ}$ Ir-Ir bonds can be labelled as X-, Y-, and Z-bonds; the Z-bond is the bridging bond along the $c$-axis and the X and Y-bonds form the zig-zag chains. In the orthorhombic structure at ambient pressure (space group: $Fddd$), the X- and Y-bonds are symmetry-equivalent. The X- and the Y-bonds are appreciably longer than the Z-bond by 3\% at ambient pressure.

With the application of pressure, the orthorhombic unit cell displays an anisotropic contraction as reported in Ref.~ \cite{beta-Li2IrO3_Veiga} (see Supplemental Materials \cite{Supplementary}). The $b$-axis lattice constant shows a stronger pressure dependence than those of the $a$-axis and the $c$-axis, which comes from the rapid contraction of X- and Y- bonds. By comparing the bond lengths at ambient pressure (Fig. 1(b)) and at a pressure of 2.6(1) GPa (Fig. 1(d)), the X- and Y-bond lengths decrease by 1.3\%, whereas that of the Z-bond actually increases by 0.5\%. As a result, the X- and Y-bond lengths and Z-bond length become much closer at 2.6 GPa than at ambient pressure.  The Ir-O bond lengths do not show any appreciable change from ambient pressure to 2.6 GPa, meaning that the change in the Ir-Ir bond lengths are controlled by the Ir-O-Ir angle. In accord with this, the Ir-O-Ir angles for X- and Y-bonds and Z-bond became much closer. (Fig. 1(c) and (d)).

\begin{table}
\caption{Refined structural parameters of $\beta$-Li$_2$IrO$_3$ in the high-pressure phase at 4.4(1) GPa. The space group is $C2/c$ (No. 15) and $Z$ = 8. The lattice parameters are $a$ = 5.8390(4) \AA, $b$ = 8.1297(5) \AA, $c$ = 9.2240(6) \AA, and $\beta$ = 106.658(4)$^{\circ}$. $g$ and $U_{\rm iso}$ denote site occupancy and the isotropic displacement parameter, respectively. $U_{\rm iso}$ was constrained to be equal across sites containing the same element during the refinement.  The refinement indices are $R_p$ = 2.69\%,  $R_{wp}$ = 2.27\% and $\chi^2$ = 2.472. The Rietveld fits to the data are available in Supplemental Materials \cite{Supplementary}.}
\begin{ruledtabular}
\begin{tabular}{ccccccc}
Atom & Site & $g$ & $x$ & $y$ & $z$ & $U_{\rm iso}$ (\AA$^2$)\\
\colrule
Li1 & 8$f$ & 1 & 0.261(4) & 0.6439(18) & 0.2437(22) & 0.0118(15)\\
Li2 & 8$f$ & 1 & 0.9345(28) & 0.6255(22) & 0.6007(15) & 0.0118(15)\\
Ir1  & 8$f$ & 1 &  0.4241(6) & 0.3857(5) & 0.0788(5) & 0.0058(4)\\
O1 & 8$f$ & 1 & 0.7262(10) & 0.3894(8) & 0.2516(7) & 0.0058(3)\\
O2 & 8$f$ & 1 & 0.9025(12) & 0.3614(6) & 0.5816(8) & 0.0058(3)\\
O3 & 8$f$ & 1 & 0.4067(12) & 0.3666(6) & 0.5835(8) & 0.0058(3)\\
\end{tabular}
\end{ruledtabular}
\end{table}

With further increase of pressure, a structural transition from the low-pressure orthorhombic ($Fddd$) to the high-pressure monoclinic ($C2/c$) structure takes place at around 3.7 GPa \cite{Supplementary}. The result of structural refinement for the high-pressure phase at 4.4(1) GPa is listed in Table 1. The hyperhoneycomb network made of edge-shared IrO$_6$ octahedra is maintained (Fig. 1(e)). The average length of X- and Y-bonds along the zig-zag chains decreases further to $\sim$2.87 \AA\;and the bridging Z-bond length increases to 3.026(6) \AA. This appears to be an extension of the anisotropic pressure dependence between X(= Y) and Z in the low-pressure phase.  A modulation of the Ir-Ir bond length in the zig-zag chains, however, makes the high-pressure phase distinct from the low-pressure phase. As seen in Fig. 1(f), the X-bond and Y-bond are no longer equivalent at 4.4 GPa. While the X-bond length is as long as 3.069(5) \AA\;close to or even longer than the increased Z-bond length, the Y-bond length becomes as short as 2.663(5) \AA, which gives rise to an alternating arrangement of the short Y-bonds and the long X-bonds along the zig-zag chains. In fact, at a distance of 2.66 \AA, the Ir-Ir Y-bond is even shorter than that seen in metallic Ir, indicating the formation of an Ir$_2$ dimer molecule in the zig-zag chains \cite{dimer_theory}.

\begin{figure}
\epsfysize=85mm
\includegraphics[width=8cm, clip]{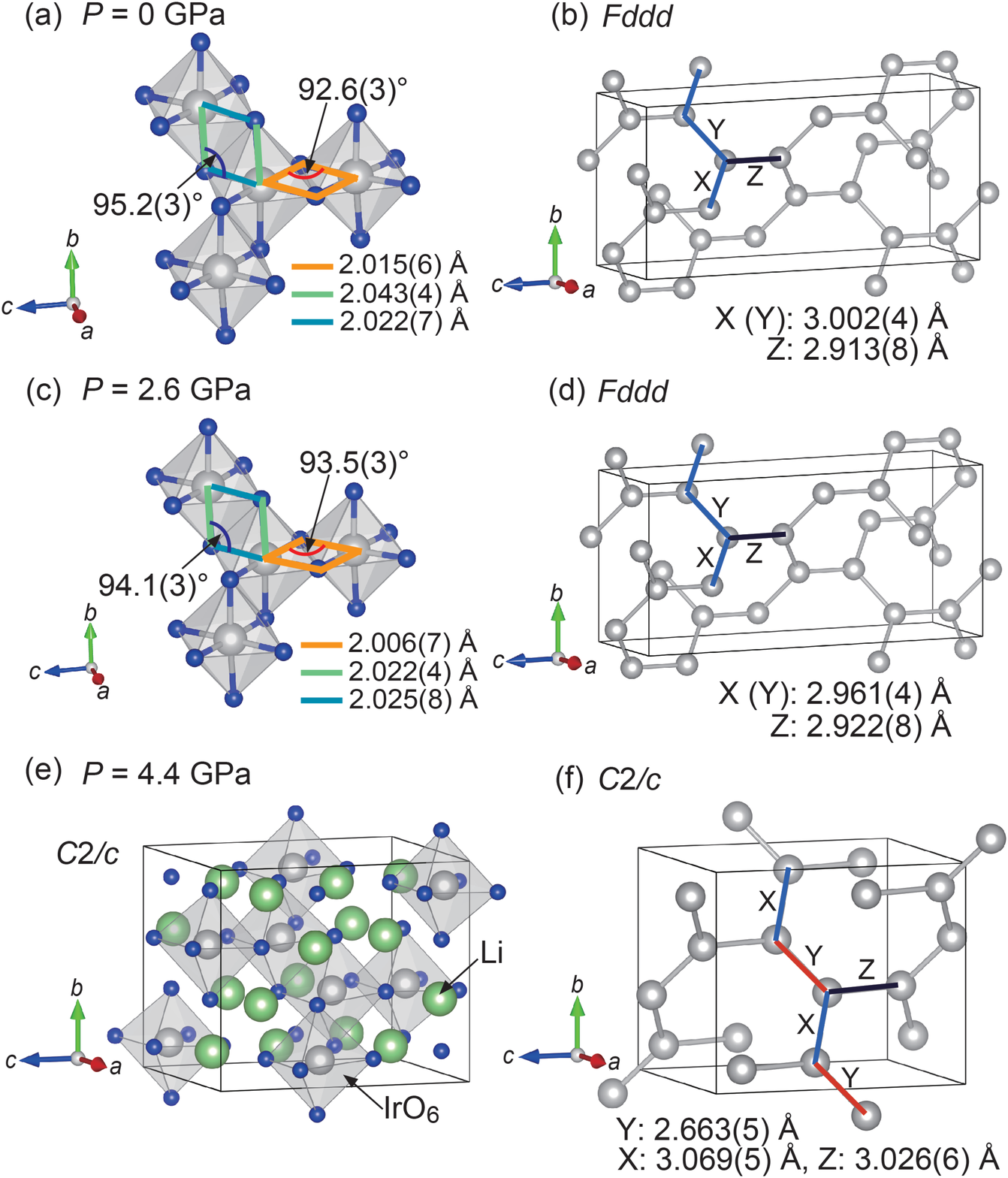}
\caption{Crystal structure of $\beta$-Li$_2$IrO$_3$. (a), (c) local structure around an IrO$_6$ octahedron at 0 and 2.6 GPa, respectively. (b), (d) Hyperhoneycomb network of Ir atoms. X, Y and Z denote the 3 types of Ir-Ir bonds. (e) Crystal structure in the high-pressure monoclinic phase at 4.4(1) GPa. (f) Ir network in the high-pressure phase. The dimerized bond is shown in red. The crystal structures are illustrated using VESTA software \cite{VESTA}.}
\label{fig1}
\end{figure}

\subsection{Collapse of $J_{\rm eff}$ = 1/2 state in the high-pressure dimerized phase}

RIXS measurements unveil the drastic reconstruction of the electronic structure associated with this dimerization. The Ir $L_3$-edge RIXS spectrum at ambient pressure, measured on a polycrystalline pellet, is displayed at the bottom of Fig. 2. In addition to the elastic scattering peak at 0 eV, there are two pronounced features; a sharp peak at around 0.7 eV and a broad peak centered at around 3.5 eV. The latter corresponds to the excitations from Ir 5$d$ $t_{2g}$ to $e_g$ manifolds. The peak at $\sim$0.7 eV represents the local excitation between the filled $J_{\rm eff}$ = 3/2 and the half-filled $J_{\rm eff}$ = 1/2 state, as observed in a number of $d^5$ iridium oxides \cite{Ishii_RIXS2011, BJ_RIXS2012, Gretarsson_RIXS2013, Liu_RIXS2012}. The peak energy represents the spin-orbit splitting between $J_{\rm eff}$ = 3/2 and $J_{\rm eff}$ = 1/2, namely 3$\lambda_{\rm SO}$/2. This evidences the dominant $J_{\rm eff}$ = 1/2 character of $t_{2g}$ holes in $\beta$-Li$_2$IrO$_3$ at ambient pressure.

RIXS spectra under pressure were collected with a single crystal loaded in a DAC. At a low pressure of 0.9(1) GPa, the spectrum is almost identical to the polycrystalline data. This supports the idea that the $d$-$d$ excitations show only a small $\bm{q}$-dependence in $\beta$-Li$_2$IrO$_3$. With increasing pressure up to 3.1(1) GPa, the 0.7 eV peak remains at the same energy but tends to be moderately broadened.  This likely suggests a pressure-induced change of electronic structure, for example, the mixing of $J_{\rm eff}$ = 3/2 and 1/2 state by enhanced hopping. We note that a pressure-induced change of electronic structure under pressure, prior to the structural transition, was inferred also by the loss of magnetic field induced ferromagnetic moment and the pronounced change of the branching ratio of the x-ray absorption spectrum above 2 GPa \cite{beta-Li2IrO3_Veiga}.

A pronounced change of RIXS spectra was found in the high-pressure dimerized phase above 4 GPa. The 0.7 eV peak is suppressed almost completely. The one at 3.5 eV is significantly broadened, more significantly on the high-energy side, but remains in the high pressure phase.  Instead of the 0.7 eV-peak, a broad continuum spreads roughly from 0.5 to 2.5 eV and a peak feature around 2.8 eV emerges. The drastic change of RIXS spectra, along with the suppression of the 0.7 eV peak, clearly points to the collapse of the spin-orbit coupling splitting of $J_{\rm eff}$ = 1/2 and $J_{\rm eff}$ = 3/2 states in the high-pressure phase. We note that the spectrum of the low pressure phase was recovered after depressurization, excluding the possibility of irreversible chemical change of the sample due to x-ray irradiation and/or pressure.

\begin{figure}
\epsfysize=75mm
\includegraphics[width=7cm, clip]{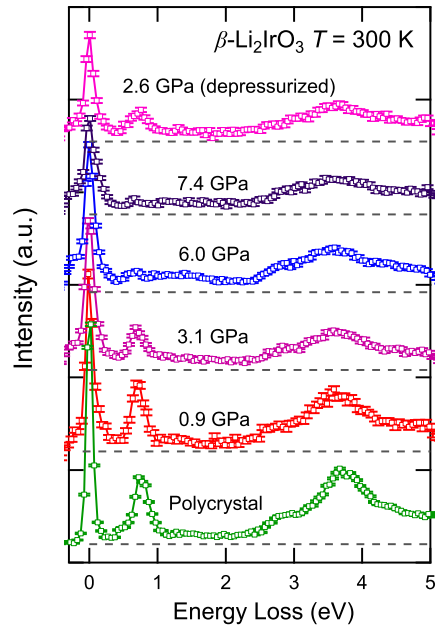}
\caption{RIXS spectra of $\beta$-Li$_2$IrO$_3$ under pressure recorded at room temperature. The data from the polycrystalline sample (bottom) were collected at ambient pressure, and the data at 2.6(1) GPa (top) after the sample was depressurized from 7.4(1) GPa. The spectra are shown with arbitrary offsets. The horizontal broken-lines represent the guide base lines, obtained by subtracting a constant background from each spectrum.}
\label{fig2}
\end{figure}

\subsection{Formation of molecular orbital in the Ir$_2$ dimers}

The electronic structure calculations were performed using the structural parameters in the ambient and the high-pressure phases, as shown in the plot of partial densities of states from the relevant orbitals in Fig. 3(a) and (b), respectively. Spin-orbit coupling was incorporated to the calculation. On-site Coulomb $U$ was not explicitly introduced, which makes the ambient pressure phase metallic. In Fig. 3(a) for the low pressure phase, the splitting of $t_{2g}$-derived bands into the $J_{\rm eff}$ = 3/2-derived bands around -1 eV and $J_{\rm eff}$ = 1/2-derived bands around the Fermi energy by spin-orbit coupling is clearly seen as in the previous calculations \cite{beta-Li2IrO3_EPL, beta-Li2IrO3_Veiga}. Appreciable mixing between $J_{\rm eff}$ = 3/2 and 1/2 can be recognized as in other iridium oxides due to the presence of trigonal distortion and/or the $J_{\rm eff}$ = 1/2 $\leftrightarrow$ 3/2 hopping between the neighboring Ir sites \cite{Y2Ir2O7_Shinaoka, Na2IrO3_Sohn}.

In stark contrast, the $d_{xy}$-, the $d_{yz}$- and the $d_{zx}$-orbitals, instead of the spin-orbital-entangled $J_{\rm eff}$ = 1/2 and 3/2 states, appear to represent the character of the bands in the high-pressure phase in Fig. 3(b). The $d_{zx}$ orbital, directed along the dimer bond (Y-bond) provides the dominant character of the two sub-bands stemmed out of $t_{2g}$ derived bands; the lowest (-1.7 eV) occupied sub-band and the highest (+ 0.7 eV) empty sub-band. The two sub-bands with predominant $d_{zx}$ character can be assigned to the bonding and the antibonding states of Ir$_2$ dimer molecules with a splitting energy of $\sim$2.4 eV. The other $t_{2g}$ orbitals, $d_{xy}$- and $d_{yz}$-orbitals mainly contribute to the sub-bands between the $d_{zx}$- bonding and antibonding sub-bands. Because of the degeneracy of $d_{xy}$- and $d_{yz}$-orbitals, strong spin-orbital-entanglement is expected for these $d_{xy}$- and $d_{yz}$-derived sub-bands, which are denoted as “entangled $xy$-$yz$” orbitals with different colors in Fig. 3(b). Since the hybridization of entangled $xy$-$yz$ orbitals between the nearest neighbor Ir atoms is much weaker than that of $d_{zx}$ orbitals, it is natural that they reside in between the bonding and antibonding orbitals of $d_{zx}$. As a result, Ir $d^5$ electrons fill up the bonding $d_{zx}$ sub-bands and the four entangled $xy$-$yz$ sub-bands. An energy gap is formed between the entangled $xy$-$yz$ sub-bands and the empty antibonding $d_{zx}$ sub-band, yielding a band insulating state. The bandwidth of occupied states increases appreciably as compared with that at ambient pressure, which is consistent with the broad feature observed in the RIXS spectra at the low-energy region up to 2.5 eV. We note that the effect of electron correlations may narrow down the bandwidth and make the RIXS peaks sharper. The $e_g$ orbitals are almost degenerate at ambient pressure, but split appreciably at 4.4 GPa due to the strong distortion of IrO$_6$ octahedra. This accounts for the broadening of the RIXS peak at 3.5 eV in the high-pressure phase.

The formation of Ir$_2$ dimers in the hyperhoneycomb lattice gives rise to bonding and antibonding molecular orbital states made of $d_{zx}$ orbitals in the bonding plane. The large bonding-antibonding splitting stabilizes a $d_{zx}$-orbital-dominant antibonding state of $t_{2g}$ holes and makes the system a band insulator, which is consistent with the negligible XMCD above 4 GPa \cite{beta-Li2IrO3, beta-Li2IrO3_Veiga}.  The emergence of the $d_{zx}$-orbital dominant state results in the collapse of the $J_{\rm eff}$ = 1/2 state. 

\begin{figure}
\epsfysize=75mm
\includegraphics[width=7cm, clip]{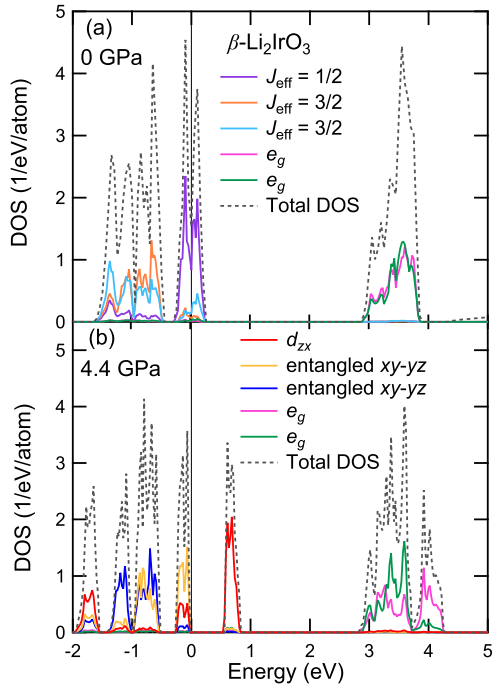}
\caption{Calculated density of states (DOS) for Ir 5$d$ states. (a) DOS for the ambient-pressure phase ($Fddd$). The Ir $t_{2g}$ orbitals are resolved into $J_{\rm eff}$ = 1/2 and 3/2 characters. (b) DOS for the high-pressure phase at 4.4 GPa ($C2/c$). The $d_{zx}$-orbital is directed along the Ir$_2$ dimer bond. The other $t_{2g}$ orbitals, $d_{xy}$ and $d_{yz}$,  are entangled by spin-orbit coupling. The total DOS includes the contributions from oxygen 2$p$ states.}
\label{fig3}
\end{figure}
 
\section{Discussion}

The present result points to a competition between the spin-orbital-entanglement and the dimerization in $\beta$-Li$_2$IrO$_3$. The former mixes up the different orbital states to create the orbital moment. The latter selects a specific orbital to gain bonding energy. It was theoretically discussed that $\beta$-Li$_2$IrO$_3$ shows an intrinsic instability towards the formation of Ir$_2$ dimers when spin-orbit coupling is neglected \cite{beta-Li2IrO3_Kee}. At ambient pressure, the spin-orbit coupling overcomes the dimer instability and the $J_{\rm eff}$ = 1/2 state is formed. By increasing the overlap of orbitals, the instability is enhanced and eventually the dimer phase shows up. Similar dimerization under pressure was recently identified in another honeycomb iridate $\alpha$-Li$_2$IrO$_3$ \cite{alpha-Li2IrO3_Hermann, alpha-Li2IrO3_Clancy}. We note that in the previous theoretical approaches, the dimerization is predicted to take place within the Z-bond \cite{beta-Li2IrO3_Kee}. In both $\alpha$-Li$_2$IrO$_3$ and $\beta$-Li$_2$IrO$_3$, the dimers are formed in the zig-zag chains made of the X- and Y-bonds rather than the Z-bond. The anisotropic lattice contraction of the zig-zag chains under pressure seems to be closely related to the preferable dimerization in the zig-zag chains, the origin of which is worthy of further exploration. The dimerization of transition-metal ions has been frequently seen in not only in honeycomb-based 5$d^5$ iridates but also in a wide variety honeycomb based 3$d$ and 4$d$ oxides and halides, including $\alpha$-TiCl$_3$ (3$d^1$) \cite{TiCl3}, $\alpha$-MoCl$_3$ (4$d^3$) \cite{MoCl3} and Li$_2$RuO$_3$ (4$d^4$) \cite{Li2RuO3} even at ambient pressure. The occurrence of dimerization only under a high pressure may reflect that the competition with the spin-orbital-entangled phase is much more significant in the 5$d^5$ iridates with spin-orbit coupling of $\sim$0.5 eV, much larger than those of the ambient pressure dimerized compounds.

The dimer transition at $P_{\rm S}$ $\sim$ 4 GPa occurs at room temperature in $\beta$-Li$_2$IrO$_3$. The disappearance of magnetic ordering in $\beta$-Li$_2$IrO$_3$ in the $T$ = 0 limit was reported to occur at $\sim$2 GPa [34, 36].  The dimerization at 4 GPa at room temperature could be suppressed to a lower pressure with decreasing temperature, and may compete with the the magnetically ordered phase around 2 GPa. The low temperature structure under pressure is worthy of further exploration to fully disclose the phase competition inherent to the honeycomb-based iridium oxides. In fact, the sister compound $\gamma$-Li$_2$IrO$_3$ shows an analogous pressure collapse of the magnetic ordering at $P_c$ = 1.5 GPa but no signature of structural dimerization up to 3.3 GPa [35].  It may be interesting to infer that the instability to dimer formation may be relevant for the breakdown of magnetic order in such 3D-based honeycomb iridates.

%\begin{figure}
%\epsfysize=70mm
%\centerline{\epsffile{Figure4.eps}}
%\caption{Magnetic susceptibilities of Sr$_2$IrO$_4$ in the high-temperature
%  paramagnetic phase. Red and blue dots show the experimental in-plane
%  ($\chi_{ab}$) and out-of-plane ($\chi_{c}$) susceptibilities, and the black
%  solid line delineates the fitting line for $\chi_{ab}$ based on Eq. (4). The
%  inset depicts the local spin frame ($\tilde{x}$, $\tilde{y}$) rotated by the
%  angle of $\pm\phi$ from the laboratory frame of ($x$, $y$). A and B
%  represent the two antiferromagnetic sublattices.}
%\label{fig4}
%\end{figure}

%\begin{figure}
%\epsfysize=50mm
%\epsfysize=60mm
%\centerline{\epsffile{Figure5.eps}}
%\caption{Interlayer couplings in nearest
%  ($J^{\prime}_{1c}$ and $J^{\prime\prime}_{1c}$) and next-nearest
%  ($J_{2c}$) planes.
%The iridium sublattice formed by anticlockwise (clockwise) rotated
%  octahedra is labeled by $\mathsf{A}$ ($\mathsf{B}$).}
%\label{fig5}
%\end{figure}

\section{Conclusion}

We studied the crystal and electronic structures of the hyperhoneycomb iridate $\beta$-Li$_2$IrO$_3$ in the high-pressure phase above 4 GPa. The high pressure phase is characterized by the formation of Ir$_2$ dimers on the zig-zag chains. The spin-orbital-entangled $J_{\rm eff}$ = 1/2 states break down, associated with the stabilization of the bonding state of the neighboring $d_{zx}$-orbitals in the dimer phase. Such competition of spin-orbital-entanglement and dimer formation are indeed widely observed in honeycomb-based iridates, and we argue it is one of the hallmarks of the physics of these materials.

\section*{Acknowledgements}

We are grateful to D. Haskel, L. S. I. Veiga, S. K. Choi, K. Kitagawa and R. Dinnebier for helpful discussions. We thank S. K. Choi, H. H. Kim, U. Engelhardt and K. Syassen for their help in the preparation of DAC experiments. We acknowledge the provision of beamtime for the PEARL experiment (Proposal No. RB1710302) to Science \& Technology Facilities Council (STFC), and the allocation of beamtime at BL11XU of SPring-8 (Proposal No. 2017-1-119-1/2017A4253, 2017-1-119-3/2017B4246, and 2017-3-078-3/2018A4260) to the National Synchrotron Radiation Research Center (NSRRC) and the Japan Synchrotron Radiation Research Institute (JASRI). This work was partly supported by the Alexander von Humboldt foundation, Japan Society for the Promotion of Science (JSPS) KAKENHI (No. JP15H05852, JP15K21717, 17H01140), and JSPS Core-to-core program “Solid-state chemistry for transition-metal oxides”

%%%%%%%%%%%%%%%%%%%%%%%%%%%%%%%%%%%%%%%%%%%%%%%%%%%%%%%%%%%%%%%%%

\end{document}